\documentclass[conference]{IEEEtran}
\IEEEoverridecommandlockouts
\usepackage{cite}
\usepackage{amsmath,amssymb,amsfonts}
\usepackage{algorithmic}
\usepackage{graphicx}
\usepackage{textcomp}
\usepackage{xcolor}
\usepackage[utf8]{inputenc}
\usepackage[ruled,vlined,linesnumbered]{algorithm2e}
\usepackage{dirtree} 
\usepackage{amsthm}
\usepackage{amsmath}
\theoremstyle{definition}
\newtheorem{definition}{Definition}[section]
\usepackage{url}
\usepackage{enumitem}
\usepackage{graphicx}
\graphicspath{ {.//} }
\usepackage{todonotes}
\usepackage{caption}
\usepackage{subcaption}
\def\BibTeX{{\rm B\kern-.05em{\sc i\kern-.025em b}\kern-.08em
    T\kern-.1667em\lower.7ex\hbox{E}\kern-.125emX}}
\begin{document}

\title{Multi-agent Path Finding in\\Continuous Environment
}

\author{\IEEEauthorblockN{Kristýna Janovská}
\IEEEauthorblockA{\textit{Faculty of Information Technology} \\
\textit{Czech Technical University in Prague}\\
Prague, Czech Republic \\
janovkri@fit.cvut.cz}
\and
\IEEEauthorblockN{Pavel Surynek}
\IEEEauthorblockA{\textit{Faculty of Information Technology} \\
\textit{Czech Technical University in Prague}\\
Prague, Czech Republic \\
pavel.surynek@fit.cvut.cz}
}

\maketitle

\begin{abstract}
We address a variant of multi-agent path finding in continuous environment (CE-MAPF), where agents move along sets of smooth curves. Collisions between agents are resolved via avoidance in the space domain. A new Continuous Environment Conflict-Based Search (CE-CBS) algorithm is proposed in this work. CE-CBS combines conflict-based search (CBS) for the high-level search framework with RRT* for low-level path planning. The CE-CBS algorithm is tested under various settings on diverse CE-MAPF instances. Experimental results show that CE-CBS is competitive w.r.t. to other algorithms that consider continuous aspect in MAPF such as MAPF with continuous time.
\end{abstract}


\section{Introduction} Recent research in path planning not only deals with discrete environments but becomes focused continuous space to reflect needs of practice. The multi-agent path finding (MAPF) is a problem in which several agents move in a shared environment without colliding and each agent must reach its goal positons. For a discrete problem, this means that two agents are not present in the same vertex at the same time step in a graph that models the environment. However, in a continuous problem that aims to realistically solve real-life scenarios such as warehouse logistics or even autonomous vehicle movement, both time and space are continuous. The agents themselves should also be represented by a rigid body. Collisions between agents should therefore mean that two agents overlap at a given time. It is also possible to take various kinematic and dynamic constraints into account which constrain the agent's movement.

The goal of this work is to create a software prototype that allows continuous movement in both time and space for multiple agents at once. Agents will move on a set of smooth curves that will create their paths. These paths will be constrained so that no agents collide at the same time. As agents are rigid circular bodies in this work, a collision will occur if two agents overlap at any time. An algorithm based on existing Continuous Conflict-Based Search \cite{ccbs} will be proposed. This algorithm works in continuous time, but the proposed modification will also work in a continuous environment. Unlike CCBS, agents will plan their paths using the RRT* algorithm \cite{RRTstar}, whose output will then be smoothed with B-spline curves. Single-agent methods for smooth path planning have already been studied. This work will build on them to provide a non-conflicting smooth multi-agent solution.

The work is sectioned as follows. Firstly, the theoretical background of this problem will be studied. This will contain the problem of continuous multi-agent path finding, the most important algorithms such as CCBS \cite{ccbs} or RRT*\cite{RRTstar}, and smoothing with B-spline curves, as well as selected related works to all of these topics will be discussed. Next, the proposed model consisting of several base parts will be explained. These parts will include a constrained single-agent path finding algorithm, a smoothing mechanism, and a multi-agent algorithm for path finding in a continuous environment called CE-CBS. In the last part, performed experiments are discussed, with several important parameters being introduced and their impact on the model's performance being studied.

    \section{Related work}
    In \cite{SurynekSwarms}, a novel solution for the MAPF$_R$ problem was proposed. The SMT-CBS$_R$ provides a makespan-optimal solution to this problem considering continuous space and time with geometric agents. It combines CBS algorithm with satisfiability modulo theory using lazy construction of incomplete encodings. That way, not all constraints are introduced in the model, which makes it consider a subset of solutions. The produced solution can then be further refined if needed. This makes it possible for the solution to be found before complete specification of the model. Unlike the original CBS, SMT-CBS$_R$ does not branch after encountering a conflict, and instead refines the propositional model with a conflict elimination disjunctive constraint (a mutex). The results of this approach are comparable to those of CCBS.
    
     In \cite{MARRTstar}, the authors proposed a multi-agent variant of the RRT * algorithm, MA-RRT *. The agents in this model moved on a discrete motion graph in Euclidean space. As the environment was discrete, the graph version of RRT*, G-RRT* was used. The main difference between the two variants is the steering technique, which in G-RRT* is done using heuristic-guided greedy search. In MA-RRT*, greedy search generates a sequence of joint actions of all agents. After each action is generated, the set of actions is checked for possible conflicts. This approach was evaluated to be more efficient than A* in joint-state space when path finding is carried out in sparse, large environments, at the cost of slight solution quality.

     The authors of \cite{optimizedRRTstar} presented ORRT-A*, an optimised version of the RRT-A* algorithm. Robots in this case move in a partially known environment.  After creating and optimizing a path with modified RRT and A* heuristic function, it is smoothed via a cubic spline. Every data point interval is represented using different cubic functions and in this way the interpolation technique smooths the path with the generation of more data points.  Experiments suggested that ORRT-A* enhances path quality in comparison to goal-biased RRT and RRT-A*.
 
The MA-RRT*FN algorithm was proposed in \cite{memoryMARRT}, which aimed to increase the efficiency of the MA-RRT * algorithm by limiting the number of nodes stored in a tree to a fixed number. To avoid reaching this limit of nodes, all nodes are evaluated, and those which are unlikely to reach the goal are removed from the path. This solution does not exceed MA-RRT* concerning path quality, although the results are close, but it is much more efficient considering memory usage.

\cite{phdVerbari} focuses on decentralized multi-agent path finding using RRT* algorithm. The author proposes Multi-RRT* algorithm, an approach aiming to find a Pareto optimal solution with using a priority order method. The problem is solved agent by agent by sorting them through a performance measure. It removes agents less involved in a collision from the problem, therefore solving each iteration in a smaller problem. Each agent computes its own trajectory with respect to solutions of other agents, which are considered as known, time-varying obstacles.

\cite{interpolatingBridges} focused on MAPF with continuous time. Agents are controlled globally and use general linear dynamics. The authors focus on navigating agents through complicated environment such as crowded areas or narrow spaces. These challenging parts are called bridges, and they represent collision-free regions of the environment with certain geometric navigation characteristics. Trajectory of each agent is computed with kinodynamic RRT*, and it moves an agent along an interpolated path in the bridge. A scheduling system is present to ensure efficient and collision-free motion.
    \section{Continuous-Environment MAPF} The problem of continuity in MAPF can be addressed from several viewpoints -- with continuous representation of time, or a continuous representation of environment. In discrete problems, time would be represented as discretised time steps, and the environment where agents are present is represented as a graph in whose vertices the agents are situated. \par
    $MAPF_R$ is a version of the MAPF problem with continuous time. Algorithms that address this problem are, for example, $CCBS, E-ICTS$ or $ECBS-CT$ \cite{ccbs, eicts, ecbsct}. Of these, only $CCBS$ provides a solution without the need for any discretization of time.

    In this work, both continuous environment and time are assumed. The agents are to find non-conflicting paths satisfying several properties. The paths are defined as smooth curves and have to satisfy certain kinematic constraints. Here, that constraint is the minimal angle constraint, setting a minimal value to angles between path segments before smoothing takes place. Agent's paths can cross, but the agents themselves cannot collide - they cannot overlap at any time. The movement of agents on their paths in time is represented by their trajectories, which are checked for conflicts. As in standard MAPF, agents start in pre-set starting position and they try to reach their goal positions.

The following definition is based on existing definitions by \cite{psNPmapf, COBRA, SeventhOrderBezier, MARRTstar} and \cite{ ccbs}.

\begin{definition}[Smooth Continuous MAPF - SC-MAPF]
    Let $\mathcal{M}$ be a metric space, and $R = \{r_1, r_2, ..., r_v\}$ a set of agents.

 SC-MAPF is the tuple $\mathcal{E}=[\mathcal{G}=(V,E),\mathcal{M}, S, G, coord, \mathcal{A}]$, where $\mathcal{G}$ represents a graph, $ \mathcal{M}$ a metric space, $S$ the start function and $G$ the goal function with $coord$ mapping every vertex in $\mathcal{G}$ to a coordinate in $\mathcal{M}$ and $\mathcal{A}$ being a finite set of possible move actions. Every edge E in $\mathcal{G}$ represents a set of smooth curves that connect two vertices in $\mathcal{G}$. A set of curves belonging to an edge $e \in E$ is denoted as $c_e$.
 
     Every action $a \in \mathcal{A}$ in MAPF$_R$ is defined by a duration $a_D$ and a motion function $a_\varphi: [0, a_D]: \rightarrow \mathcal{M}$ that maps time to metric space. This motion function corresponds to a smooth curve. $a_\varphi (t)$ is the coordinate of an agent in the metric space $\mathcal{M}$ at a time $t$ while executing an action $a$. 

     The agent body is defined as a shape in metric space.
    
    The initial configuration of agents is defined by a simple function $S_R: R \rightarrow V$, $S_R(r) \neq S_R(s)$ for each $ r, s \in R, r \neq s$. \par
    
The goal configuration of agents is a simple function $G_R: R \rightarrow V $, $G_R(r) \neq G_R(s)$ for each  $r, s \in R, r \neq s$. \par

A problem of smooth continuous multi-agent path finding is a task of finding a set of smooth curves - a sequence of actions, each choosing a curve $c$ from a set of curves $c_e$ belonging to a respective edge $e$, for each agent so that no agents collide.

A conflict is defined as a 5-tuple $\mathcal{C} = [\alpha_i, a(\alpha_i), \alpha_j, a(\alpha_j), t]$, where $\alpha$ is an agent, $a(\alpha)$ is the action agent $\alpha$ performs at the time the conflict occurs and $t$ is the beginning of the time interval when the conflict occurs. This means that two agents overlap. Solving a conflict means forbidding the agents to move on certain curves which would traverse through the conflicting position at a conflicting time.

The solution of this problem is a set of actions of a set of agents satisfying the following constraints:

\begin{enumerate}[label=(\roman*)]
\item $S_R = G_R$; meaning all agents reach their goal positions.
\item An agent moves at a set speed.
\item An agent's trajectory is a smooth curve, which satisfies given kinematic constraints.
\item A sequence of actions can be valid only if no two agent bodies overlap at any time.
\end{enumerate}
\end{definition}

In this work, agent bodies are defined as circles with fixed radius.
    
    \section{Background}
    \paragraph{Conflict-Based Search (CBS) and Continuous Time Conflict-Based Search (CCBS)} The CBS algorithm is a cost-optimal MAPF algorithm. The goal of this algorithm is to find a set of paths for a set of agents, so that no agents collide at any step of their paths, if a solution exists, since it is a solution-complete algorithm \cite{ccbs}. The objective is to find a conflict-free plan that minimises the makespan, which is the maximum of agents' path costs. CBS is a discrete algorithm; the environment is represented by a graph, and time is discretised into time steps. In every time step, an agent may perform one action, either a movement into a neighbouring vertex or a waiting action, that is, staying in its current vertex. 

This algorithm consists of two levels; the lower level computes a path for each agent with the help of a single-agent path finding algorithm such as A*. This path must satisfy the constraints provided by the higher level of CBS.
A constraint arises from collision solving, which takes place at the higher level. A constraint includes an agent $a$, a vertex $v$, and a time step $t$, which means that the specific agent $a$ cannot be present in $v$ at step $t$.
The higher level of CBS first computes paths for all agents and then checks each step of these paths for collisions. Collision is a situation where two or more agents are present in a vertex at a time.

The CCBS algorithm is analogous to CBS with the exception being the definition of a conflict and its resolution. In CCBS, the objective is also to minimise the makespan, defined as a maximum of costs of individual agents' paths. It is also a solution-complete algorithm, finding a solution if it exists, while unable to detect if a solution does not exist \cite{ccbs}.

In CBS, a conflict is typically represented as a tuple of an agent, a vertex, and a time step. As CCBS operates with continuous time, this is no longer viable. Instead of a vertex and a time step, CCBS takes into account an action of an agent $l$ in a calculated $unsafe$ $interval$ $[t_l , t_l^u)$. A CCBS conflict is defined using pairs of timed actions $(a, t)$. Timed action means that an action $a$ is executed from time $t$. A conflict of two agents $i$ and $j$ occurs if they execute their respective timed actions and collide. Therefore, a CCBS conflict is defined as a tuple $(i, j, (a_i, t_i), (a_j, t_j))$.

An unsafe interval of a timed action $(a_i, t_i)$ is defined with respect to another timed action $(a_j, t_j)$ as the maximal continuous time interval starting from $t_i$, in which if agent $i$ would perform the action $a_i$, it would have a conflict with timed action $(a_j, t_j)$.
\cite{ccbs}

    \paragraph{Rapidly Exploring Random Trees *} A classic algorithm for path planning in a continuous metric space is the RRT algorithm (Rapidly-Exploring Random Trees) \cite{RRT98}. A rapidly expanding random tree is constructed with iterative expansion by randomly sampling points that do not collide with an obstacle and connecting them to the nearest point in the tree. Expansions are biased towards unexplored parts of the search space. An RRT is probabilistically complete \cite{RRT98}, but the constructed paths are not guaranteed to be optimal. An improved version, RRT*, aimed to solve this problem by introducing the A* cost function into the original algorithm. Thus, the algorithm makes paths converge to optimal solutions \cite{RRTstar}.

    \paragraph{Path smoothing with B-spline Curves} In a real-life scenario, where an agent would be a vehicle, the vehicle would be sensitive for example to changes in acceleration, which would have to occur if it were to perform a sudden change of direction, such as a sharp turn. This could have an impact on the time that it spends on that path. Because of this, the differences in the different closely laid segments of the path should be limited \cite{curvesRRT, DTSP, faiglDTSP}. A possible solution is to introduce a smoothing mechanism, therefore making the path a smooth curve. The curvature of the path can then be constrained to prevent sudden turning changes. Because of these real-life implications, it is advantageous to work not only with continuous time, but with continuous environment, too, as it can introduce movement on smooth curves and therefore take various kinematic or dynamic constraints into account.

A path generated by RRT* can contain sharp turns when going from one path segment to another. To improve the quality of a path, many smoothing mechanisms were proposed, starting with \cite{1957}. This work uses smoothing via B-spline curves, so as to be able to represent the complex nature a path in an environment with many obstacles can have. A B-spline curve is fitted onto a path with concrete path points being curve control points, and additional points are then interpolated laying on the smooth curve.

 A B-spline curve is a piecewise polynomial curve. A spline is a piecewise polynomial of degree $n$ with segments $C^{n-1}$. It can be thought of as a method to define a sequence of degree $n$ Bézier curves that automatically join with $C^{n-1}$, regardless of the placement of the control point. It can be used in problems too complex to be modelled with a Bézier curve. \cite{computer_aided_design}
 
 \begin{definition}[B-spline basis function]
     Let $U=\{u_0, ..., u_m\}$ be a non-decreasing sequence of real numbers. The $u_i$ are called knots, and $U$ is the knot vector. The $i$-th B-spline basis function of $p$ degree, denoted as $N_{i,p}(u)$ is defined as follows \cite{nurbs}:
     \begin{align*}
       N_{i,0}(u)=  &\begin{cases} 
      1 & u_i\leq u < u_{i+1} \\
      0 & otherwise
   \end{cases}\\
   N_{i,p}(u) = &\frac{u-u_i}{u_{i+p}-u_i}N_{i, p-1}(u)\\
                &+\frac{u_{i+p+1} - u}{u_{i+p+1}-u_{i+1}}N_{i+1, p-1}(u)
   \end{align*}
\end{definition} 
A B-spline curve is then defined as a linear combination of control points and B-spline basis functions.
 \begin{definition}[B-spline curve]
 A p-th degree B-spline curve is defined as:
 \begin{align*}
    C(u) = \sum_{i=0}^{n} N_{i,p} (u)P_i
 \end{align*}
 where $a \leq u \leq b$, $P_i$ are control points, and $\{N_{i,p}(u)\}$ are p-th degree B-spline basis functions defined on non-periodic knot vector $U = \{\underbrace{a, ..., a}_{p+1}, u_{p+1}, ...,$ $ u_{m-p-1}, \underbrace{b, ..., b}_{p+1} \}$. \cite{nurbs}
 \end{definition} 
    \section{Proposed Model} This works builds on the Continuous Conflict-Based Search algorithm proposed by \cite{ccbs}. Here, a modification called Continuous Environment Conflict-Based Search (CE-CBS) is proposed. Its aim is to minimise the necessary discretization of the environments. It operates in a continuous environment with continuous time. 
    
    Each agent in this multi-agent simulation performs path finding via the RRT* algorithm proposed by \cite{RRTstar}, which is modified to both solve constraints provided by CE-CBS and provide a path which is then smoothed using B-spline curves. The goal is to find a conflict-free solution while minimising the criterion $sum$ $of$ $costs$, which is the total sum of the paths of all agents.
    Agents in this model are defined as circular bodies with fixed radius and speed, both of which are input parameters. They are modelled as goal-based agents \cite{russel2010} moving in a known environment. These agents aim to plan a set of paths from start positions to goal positions while avoiding both static obstacles, which are represented here as rectangles of varying sizes, and dynamic obstacles, which are constraints set by CE-CBS to avoid collisions between agents.

    In the base design of this work, agents are set to find smooth paths in a way that they can move continuously without ever stopping and waiting before reaching their goal position.

As with CCBS, the cost of a solution is the sum of the costs of each agent's individual path. \cite{improvedCCBS} The length $\rho$ of a path is measured as the sum of distances between neighbouring points on a path. The metric used is the Euclidean distance:

 The goal of the algorithm is to minimise this sum of costs while keeping the plan conflict-free.
 \paragraph{Redefining Conflicts}
 A conflict in this work is defined as a 5-tuple $C = (a_i, t_{a_{i}}, a_{i+1}, t_{a_{i+1}}, p_{i},$ $ p_{i+1})$, where $a_{i}, a_{i+1}$ are two conflicting agents, $t_{a_{i}}, t_{a_{i+1}}$ are the respective times of the agent to reach the conflict position and $p_{i}, p_{i+1}$ are the conflict coordinates for each agent.

Whether two agents are in a conflict is determined in two steps. For each pair of agents, it is checked if their planned trajectories intersect. If so, a time is computed that represents the moment when the two agents reach the intersection. This is done with the basic formula of $ t = \frac{s}{v}$, where $t$ is the resulting time, $s$ is the length of the agent's path from the start to the intersection point, measured as a sum of Euclidean distances for each path segment, and $v$ is a constant speed of an agent.

It is also necessary to account for agent bodies, for which checking if paths intersect does not suffice. It is therefore also checked if the distance between every pair of path segments from different paths is closer to each other than the agent radius. If so, the points at which those segments are closest are tested to see at which time agents arrive at these positions. If these times fall into an unsafe interval of another agent, a conflict is present.

An unsafe interval is created from these time variables, which is necessary to take the agent mass into account. If the time when agent $a_1$ reaches the intersection point falls into the unsafe interval of agent $a_2$ or vice versa, a conflict is present.

 \begin{algorithm}[h]
\SetKwInOut{Input}{Input}
\SetKwInOut{Output}{Output}
\Input{paths $P$, agents $A$}
\Output{conflict $c$}
$C \leftarrow \emptyset$\;
\For{$a_i \in A$}{
    \For{$a_{i+1} \in A$}{
        $crossing, positions \leftarrow$ $ $CrossingCloseTraj$(P_{a_i}, P_{a_{i+1}})$\;
        \If{$crossing$}{
        \For{$pos_{i}, pos_{i+1} \in positions$}{
            $t_{a_{i}} \leftarrow $TimeToReach$(P_{a_i},$ $ pos_{i})$\;
            $t_{a_{i+1}} \leftarrow $TimeToReach$(P_{a_{i+1}},$ $ pos_{i+1})$\;
            \If{$TimesTooClose(t_{a_{i}}, t_{a_{i+1}})$}{
                $C \leftarrow C$ $\cup $ Conflict$(a_i, t_{a_{i}},$ $ a_{i+1}, t_{a_{i+1}}, pos_{i}, pos_{i+1})$\;
            }
        }
        }
    }
}
\Return SelectFirst(C)\;

\caption{Get first conflict}
\label{alg:first-conflict}
\end{algorithm}

The conflict which is returned by the $GetFirstConflict$ function as seen in algorithm \ref{alg:first-conflict} is that in which one of the agents' time is the lowest from all the times of all conflicts.

\paragraph{Lower level of CE-CBS} The lower level of CE-CBS is performed by RRT* enhanced by several steps from the original algorithm proposed by \cite{RRTstar}. The lower level operates in two basic steps: path planning and interpolation. Each part has to be checked for validity to ensure the resulting path being as conflict-free and obstacle-free as possible, and iterates until a valid solution is found.

The first step is the RRT* path planning, which in itself validates the path as described below. If the path found is not valid, meaning not only that it is conflict-free and obstacle-free, but that all angles between neighbouring path segments are larger than set minimal degree constraint $\alpha$, RRT* is re-planned with a lower value of maximum nodes to sample. If a valid solution is found, interpolation with a B-spline curve can take place. The path resulting from interpolation is then validated in the same manner as the path resulting from RRT*, and interpolation can also be done multiple times to try and find a sufficient path without the need to re-planning the entire path by RRT*. Experiments showed it is necessary to be able to dynamically set the lower bound of maximal number of nodes expanded by RRT* $\eta_{max}$, so that the entire environment can be searched even if this value is lowered. To ensure that RRT* does not return a solution which is either incomplete or breaking an angle constraint, iterating continues after reaching the lower bound, but it does not lower $\eta_{max}$ further.

\paragraph{CE-CBS and kinematic constraint satisfaction}
The RRT* algorithm takes a set of constraints on input. A constraint is defined as a 4-tuple $c=(a, p, unsafe_{from}, unsafe_{to})$, where $a$ specifies an agent to which that constraint applies, $p$ is a particular coordinate of a conflict and $(unsafe_{from}, unsafe_{to})$ denotes the unsafe time interval. If an agent crossed $p$ in a time within the unsafe interval, it would mean breaking that constraint.

While path planning, RRT* has to take two types of obstacles into account -- both static predefined rectangular obstacles which remain obstacles throughout the whole course of the planning, and constraints provided by CE-CBS, which serve as dynamic obstacles and only present a threat in a certain time frame specified by the unsafe interval.

If a path segment is to be added to the RRT* tree, it first needs to be checked for validity, so that the line segment formed by adding a new point does not intersect an obstacle or break a constraint.

 \begin{algorithm}[h]
\SetKwInOut{Input}{Input}
\SetKwInOut{Output}{Output}
\Input{path $p$, obstacles $O$, constraints $C$}
\Output{True or False}
$segment \leftarrow (p_0)$\;
\For{$i$ in $length(p) - 1$}{
$segment \leftarrow segment + p_{i+1}$\;
$line \leftarrow p_i, p_{i+1}$\;
\If{not ObstacleFree$(line, O)$}{
    \Return False\;
}
$cost \leftarrow $PathCost$(segment)$\;
\If{not ConstraintFree$(line, cost, C)$}{
    \Return False\;
}
}
\Return True\;

\caption{Validate}
\label{alg:validate}
\end{algorithm}

The algorithm \ref{alg:validate} shows the function $validate$. This function checks if any line segment defined by two neighbouring points intersects an obstacle or a constraint. Intersecting a constraint means that the agent would arrive at a given position in a time specified by the unsafe interval of the constraint.

A sufficient path must satisfy the given kinematic constraints. In this work, angles between path segments (pre-smoothing) are constrained to be bigger than 90 degrees to ensure that a simulated robot would be able to preform the turn. This constraint number was preliminarily chosen so as to eliminate acute angles from paths and can be changed in further experiments.

Using the cosine formula for the dot product, the angle $\theta$ between two line segments represented by two vectors $v, w$ can be written as:

 \begin{align*}
   \theta = cos^{-1}(\frac{v \cdot w}{||v|| \cdot ||w||})
 \end{align*}

 When RRT* planning starts, it is given a constraint for the maximum number of nodes it is able to sample. This number is provided in the input. RRT* is the run with this number. Preliminary experiments have shown that in some edge cases where an obstacle is narrow and a sharp turn is path length optimal, a path with an acute angle will be generated.

\begin{figure}
        \centering
        \includegraphics[width=0.5\linewidth]{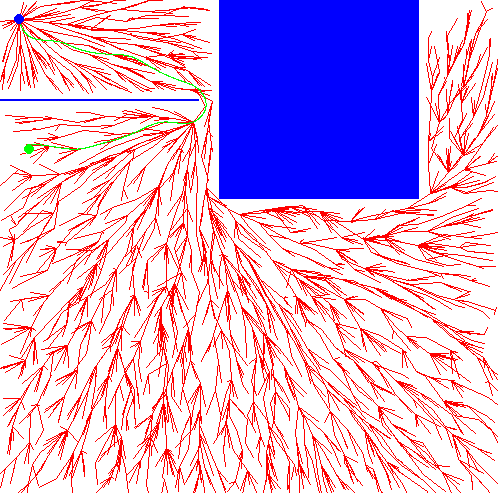}
        \caption{A smooth path avoiding a narrow obstacle successfully. Resulting path from blue circle starting position to green circle goal position is highlighted green. In red is shown the sampled RRT* tree.}
        \label{fig:curvature-ok}
    \end{figure}

 A path segment can be modified in a number of ways to correct this, as described below, but this approach may sometimes not suffice. In that case, a new iteration of RRT* is run with a lower number of maximum node number. This iteration may opt for a less optional path, but this setback can be beneficial in keeping all angles larger than the limit. If one re-planning does not suffice, more iterations follow, each with a lower maximum node parameter value, until an obstacle and conflict-free path has been found or a minimal value of this parameter is hit, if ending the search this way is allowed. Figure \ref{fig:curvature-ok} shows a path successfully navigating in a problematic environment.
 When a path is constructed by RRT*, it is validated before being passed on to be smoothed. This validation checks all angles between pairs of neighbouring path segments and tries to correct them if they are too small. If an acute angle is found, the algorithm first tries to eliminate the vertex which is shared by the two path segments and connect the other two vertices instead. If a vertex can be eliminated, it is not added to the new path $q$. This will not only shorten the original path but will account for increasing the previously acute angle to 180 degrees. This is only possible if there are no obstacles or constraints preventing connecting the other two path segments.

If a shared vertex cannot be eliminated, the algorithm then tries to move it closer to the other two vertices to increase the angle size. The point moves along a perpendicular to line segment formed by the other two vertices. The final position must be as close to the intersection point of these lines as possible without colliding with an obstacle or violating any constraints.

If neither of these approaches enlarges the angle enough, RRT* runs again with a smaller value of the maximum node value to find a possibly less optimal path that may satisfy the constraint.

\paragraph{Path smoothing}

As with the path planning itself, smoothing takes several steps. The interpolation itself is performed first. Given path points resulting from RRT*, a B-spline curve representation is found and a new point is interpolated for at least each unit of path length. Although a smooth path is found at this stage, it still needs to be validated to check if this new representation satisfies all constraints and avoids all obstacles. If it does not succeed, the interpolation may then be tried again with a larger value of the smoothing parameter $s$.

If no path is found, even then, the interpolation ends with a fail. If RRT* did not previously plan its path with a borderline low value of maximum node number $\eta_{max}$, it can be replanned with a smaller value and interpolated again. The algorithm continues to search for a valid solution even after decreasing $\eta_{max}$ to the minimum allowed number $\eta_{min}$. The search continues with $\eta_{max} = \eta_{min}$ until a valid solution is found.

\paragraph{Algorithm properties} The lower level of CE-CBS is a RRT-based algorithm, and thus is probabilistically complete. The whole model is though solution-complete, as it both does not stop until a valid solution is found for each individual path in lower level and a conflict-free solution is found in the higher level. Therefore the algorithm iterates until it finds a valid solution.

    \section{Experiments} Experiments were made to determine the model's performance under different setting of its parameters and different types of scenarios. Some of these experiments are discussed here.
    
    The performance was tested with respect to various parameters including: $\eta_{max}$ -- the maximal number of nodes RRT* can sample, $\alpha$ -- the minimal angle value between two path segments, or $r$ -- agent radius.

    \paragraph{Maximal number of RRT* nodes $\eta_{max}$} Parameter $\eta_{max}$ is essential not only for the computational time of the algorithm, but most importantly for increasing the path quality. The more RRT* is allowed to search the environment, the more it optimises its paths and approaches convergence towards the optimal solution. As running RRT* without a constrained value of $\eta_{max}$ could lead to potentially infinite computational time, it is necessary trade-off the computational time and path quality. Three agents were used in this experiment and their starts and goals were arranged in a star-shaped pattern. 8 values of $\eta_{max}$ are tested, with 10 iterations per value.

  \begin{figure}[h]
    \centering
    \begin{minipage}[b]{\linewidth}
        \centering
        \includegraphics[width=.8\linewidth]{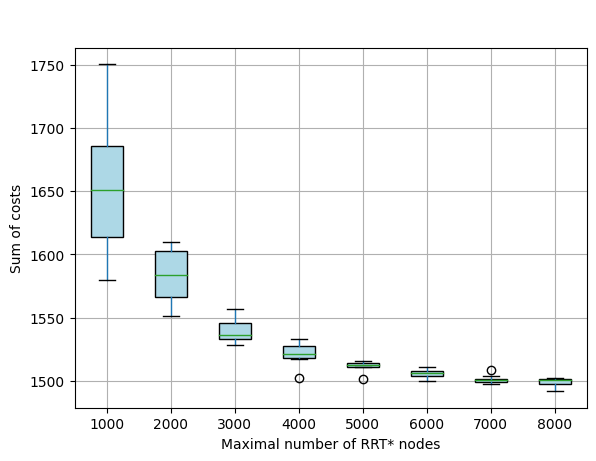}
        \caption{Figure depicting results of experiment showing relationship between average path length per agent and $\eta_{max}$.}
        \label{fig:experiment4}
    \end{minipage}
    \hspace{0.05\linewidth} 
    \begin{minipage}[b]{\linewidth}
        \centering
        \includegraphics[width=.8\linewidth]{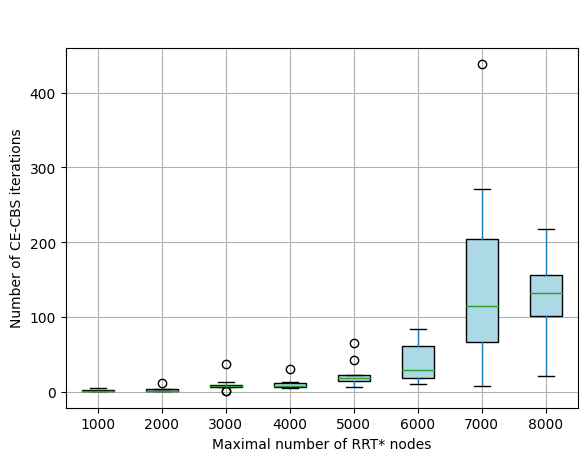}
        \caption{Figure showing how the number of CE-CBS iterations changed based on different $\eta_{max}$.}
        \label{fig:experiment4_2}
    \end{minipage}
    \label{fig:etamax}
\end{figure}

\begin{figure}[h]
    \centering
    \begin{minipage}[b]{0.45\linewidth}
        \centering
        \includegraphics[width=\linewidth]{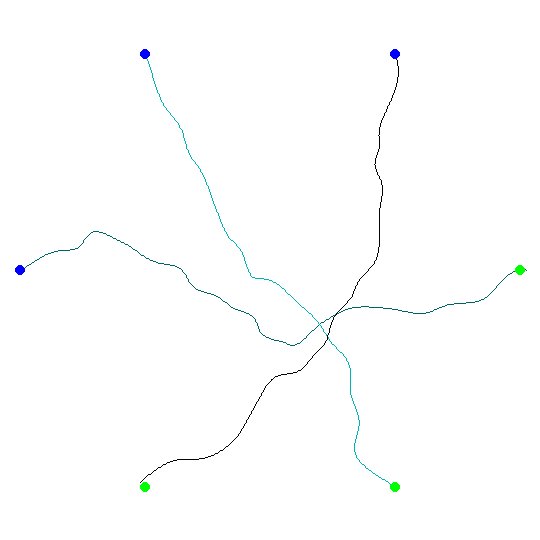}
        \caption{$\eta_{max}=1000$}
        \label{fig:obrazek1}
    \end{minipage}
    \hspace{0.05\linewidth} 
    \begin{minipage}[b]{0.45\linewidth}
        \centering
        \includegraphics[width=\linewidth]{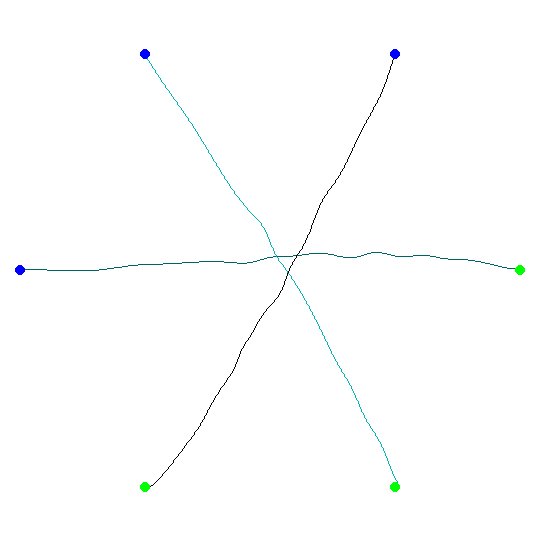}
        \caption{$\eta_{max}=6000$}
        \label{fig:obrazek2}
    \end{minipage}
    \caption{Comparison between results of runs with differently set parameter $\eta_{max}$ -- maximal number of samplet RRT* nodes.}
    \label{fig:exp4_maps}
\end{figure}

    Figure \ref{fig:experiment4} shows results for the sum of costs (y axis) that decreases with increasing $\eta_{max}$. The scatter between values also decreases with increasing $\eta_{max}$. These changes become less apparent after $\eta_{max} \geq 5000$. The opposite appears to be true for the number of CE-CBS iterations, which means that the number of CE-CBS nodes expanded during the course of the algorithm. The number of CE-CBS iterations as a function of $\eta_{max}$ is shown in Figure \ref{fig:experiment4_2}. As $\eta_{max}$ increases, so does the number of iterations, also with higher scatter. This becomes more apparent with $\eta_{max} \geq 5000$, with a large jump in both the values reached and the scatter in $\eta_{max} = 7000$. Examples of two outputs for two values of $\eta_{max}$ are shown in Figure \ref{fig:exp4_maps}.

    As higher values allow RRT* to explore the environment more, it is able to provide a path closer to the optimal one. With paths approaching the optimum however more often come situations where collisions arise, as the setting of this experiment suggests. With lower values of $\eta_{max}$, agents did not collide as often because of frequent detours they take. With more conflicts appearing, the space becomes more constrained and allows RRT* to also explore non-constrained parts more.

    \paragraph{Varying agent radius $r$} This experiment visualises a scenario in which two agents would optimally share a path, a starting node of one agent being the goal node of the second agent and vice versa. As agents are about to collide in the middle of their paths, they would have to navigate to avoid this area, but in a way which does not create another area of conflict, meaning the two agents would not be able to dodge in the same direction. The method takes into account agents' bodies and plan a path so that agents' bodies don not overlap at any time. This experiment studies different sizes of agent bodies to show how paths change with respect to this parameter.
    
\begin{figure}[h]
    \centering
    \begin{minipage}[b]{0.48\linewidth}
        \centering
        \includegraphics[width=\linewidth]{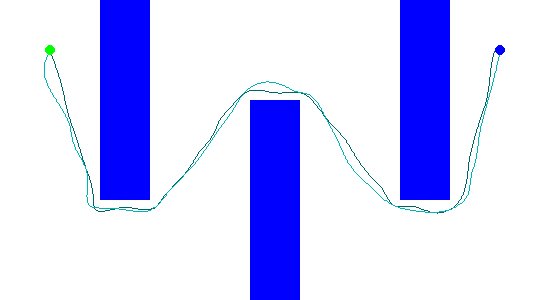}
        \caption{$r=5$}
        \label{fig:obrazek1}
    \end{minipage}
    \begin{minipage}[b]{0.48\linewidth}
        \centering
        \includegraphics[width=\linewidth]{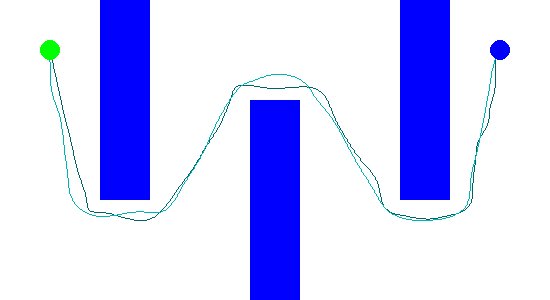}
        \caption{$r=10$}
        \label{fig:obrazek2}
    \end{minipage}
    \caption{Comparison between results of runs with differently set parameter $r$ -- agent radius.}
    \label{fig:exr}
\end{figure}

    Increasing sizes of agent bodies represented by the radius parameter $r$ were tested. Ten iterations were performed for each tested value. Several obstacles were present in the environment and agents were placed in a way which made starting position of one agent the goal position of the other agent. This was to make them plan a path that avoids overlapping with the other agent body. Visualization of two case results is presented in Figure \ref{fig:exr} 

\begin{figure}
    \centering
    \includegraphics[width=.8\linewidth]{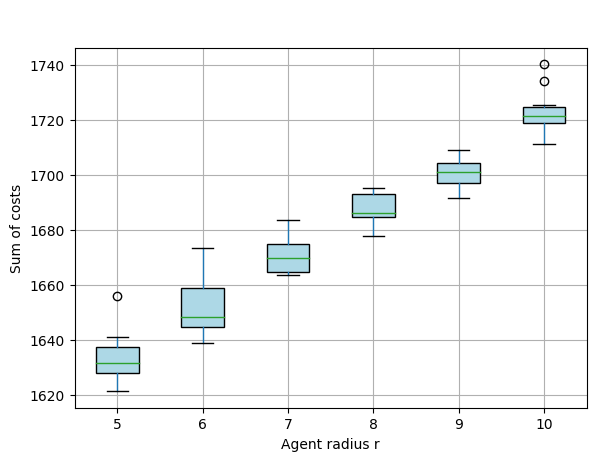}
    \caption{Result of experiment concerning agent radius $r$.}
    \label{fig:graphr}
\end{figure}

    As a result, the sum of costs increased with increasing $r$ without any exceptions as shown in Figure \ref{fig:graphr}. As the radius $r$ increased, the method planned planned longer detours for agents around the part of the environment where they would meet to accommodate their sizes. The slight decrease in the scatter between results of the same parameter values could be attributed to the fact that agents have larger bodies and must maintain a greater distance from static obstacles and set larger unsafe intervals, thus pruning the search space more with increasing $r$. 
    
    \paragraph{Comparison of CBS with CE-CBS}

\begin{figure}[ht]
    \centering
    \begin{minipage}[b]{0.45\linewidth}
        \centering
        \includegraphics[width=\linewidth]{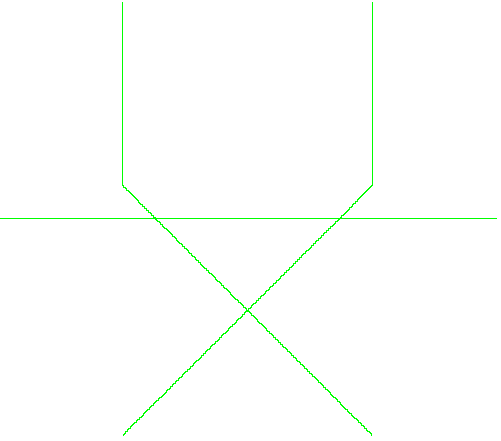}
        \label{fig:obrazek1}
    \end{minipage}
    \hspace{0.05\linewidth} 
    \begin{minipage}[b]{0.45\linewidth}
        \centering
        \includegraphics[width=\linewidth]{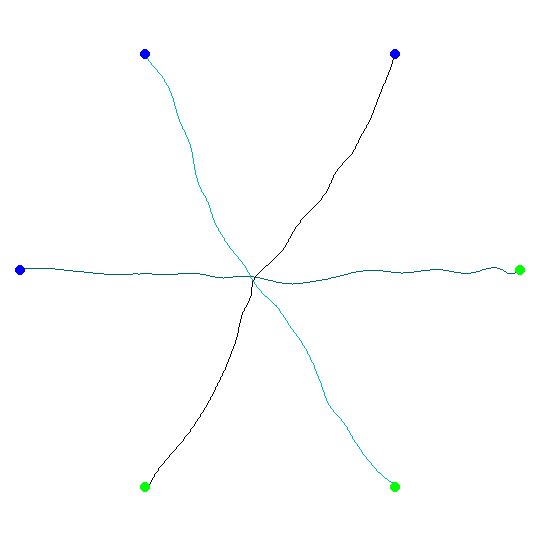}
        \label{fig:obrazek2}
    \end{minipage}
    \caption{Paths planned with discrete CBS (left) and continuous CE-CBS (right).}
    \label{fig:comparison}
\end{figure}

    CE-CBS was compared to standard CBS algorithm with A* as its low level. As CBS is a discrete algorithm, a simplified environment was used. A continuous environment of size $540 \times 540$ units was transformed into a 2D grid of $540 \times 540$ cells. Traversing one cell is understood as traversing a space of length 1. A* of this discrete CBS also used the Euclidean metric. Three agents were used for this comparison. In the discrete version, the agents' body sizes were not taken into account. The resulting paths are shown in Figure \ref{fig:comparison}. The sum of costs resulting from discrete CBS reached 1866. The average of sum of costs over 10 iterations of CE-CBS was 1522,5. This shows a measurable advantage of CE-CBS, which is able to plan shorter paths due to lifting grid movement requirements.

\begin{table}[h!]
    \centering
    \begin{tabular}{|c|c|c|c|}
\hline
\multicolumn{1}{||c|}{Map} & \multicolumn{1}{|c|}{Agents} & \multicolumn{1}{|c|}{SMT-CBS} & \multicolumn{1}{|c||}{CE-CBS} \\
\hline\hline
 $grid\_04x04\_n2$ & 2 & 4.828 & 4.46 \\
\hline
 $grid\_04x04\_n2$ & 3 & 6.243 & 5.88 \\
\hline
 $grid\_04x04\_n2$ & 4 & 9.657 & 9.04 \\
\hline
$grid\_04x04\_n2\_o$ & 2 & 7.657 & 6.24 \\
\hline
$grid\_04x04\_n2\_o$ & 3 & 9.071 & 7.70 \\
\hline
\end{tabular}
    \caption{Table comparing sums of costs between SMT-CBS and CE-CBS on two maps.}
    \label{tab:table}
\end{table}

    \begin{figure}[h]
    \centering
    \begin{minipage}[b]{0.45\linewidth}
        \centering
        \includegraphics[width=\linewidth]{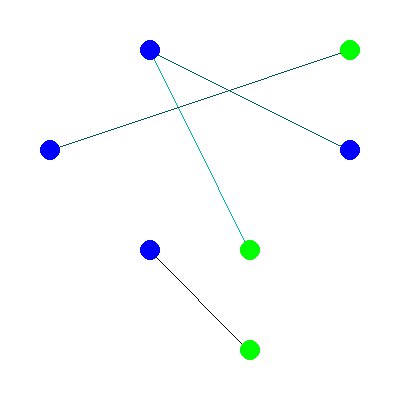}
        \caption{CE-CBS on map $4\times4$ grid with 4 agents starting at blue positions with green goal positions.}
        \label{fig:ce-cbs-a}
    \end{minipage}
    \hspace{0.05\linewidth} 
    \begin{minipage}[b]{0.45\linewidth}
        \centering
        \includegraphics[width=\linewidth]{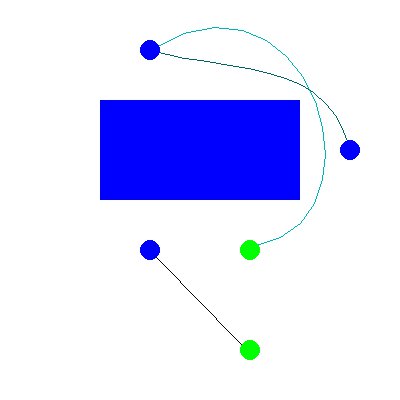}
        \caption{CE-CBS on map $4\times4$ grid with one rectangular obstacle blocking two cells and three agents.}
        \label{fig:ce-cbs-b}
    \end{minipage}
    \label{fig:ce-cbs-c}
\end{figure}

CE-CBS was also compared to SMT-CBS algorithm \cite{SurynekSwarms}. Map used for comparison was $4 \times 4$ grid from SMT-CBS dataset containing a 16 vertex grid map where diagonal movements of agents are possible. This map was also modified by adding an obstacle, this formed map $4 \times 4$ grid. Example of these maps, which were made into continuous environment for CE-CBS are shown in figures \ref{fig:ce-cbs-a} and \ref{fig:ce-cbs-b}. Sum of costs was compared in scenarios with 2 -- 4 agents. These results are shown in Table \ref{tab:table}. In all of the studied cases, CE-CBS was able to find a shorter non-conflicting path thanks to relaxed movement constraints. Although SMT-CBS found a solution in all of studied cases, it was not able to solve timed conflicts in a $4 \times 4$ grid with an obstacle. CE-CBS found a non-conflicting solution in all of studied cases. It is therefore possible that CE-CBS may perform more efficiently than SMT-CBS in real-life scenarios.

    \section{Discussion} The experiments have shown that the model can adapt to several different types of environment and that the parameters of the model can be adjusted to solve a specific scenario. It was shown that of the parameters studied, the parameter of the maximum number of RRT * nodes $\eta_{max}$ had the highest impact on the quality of the solution. The computational time corresponds to the number of CE-CBS iterations, so when more conflicts arise while planning, the computational time is higher, as RRT* is re-planned several times. The computational time is also influenced by $\eta_{max}$. It was shown that in cases with a higher number of collisions, $\eta_{max}$ can be balanced to some extent, as collisions prune the search space and allow RRT* to search more of the unconstrained part of the environment.

Based on the results of the experiments, it seems that agents are able to navigate both narrow hallways and spaces where multiple conflicts can occur at once without colliding, while still searching for the shortest path. The agents can move on smooth curves while satisfying various angle constraints and avoiding both other agents and static obstacles.

\section{Conclusion} The goal of this work was to propose a method for collision-free multi-agent path finding in continuous environment. A Continuous-Environment Conflict-Based Search (CE-CBS) algorithm was proposed, which is based on existing CCBS \cite{ccbs} and RRT* \cite{RRTstar} algorithms while also using a smoothing mechanism. Agents in this model communicate via CE-CBS and plan conflict-free paths. Individual paths are planned using RRT*, which is modified to take CE-CBS constraints and kinematic constraints into the account. The output of this algorithm is then smoothed with B-spline curves.

Future work should entail extending the model to contain a wider range of kinematic constraints.

\section{Acknowledgements} This research at the Czech Technical University in
Prague has been supported by GA\v{C}R - the Czech
Science Foundation, grant registration number 22-
31346S and the Czech Technical University in Prague Students Grant Project number SGS23/210/OHK3/3T/18.

\bibliographystyle{apalike}
\bibliography{ref}

\begin{thebibliography}{}

\bibitem[Andreychuk et~al., 2021]{improvedCCBS}
Andreychuk, A., Yakovlev, K., Boyarski, E., and Stern, R. (2021).
\newblock Improving continuous-time conflict based search.

\bibitem[Andreychuk et~al., 2022]{ccbs}
Andreychuk, A., Yakovlev, K., Surynek, P., Atzmon, D., and Stern, R. (2022).
\newblock Multi-agent pathfinding with continuous time.
\newblock {\em Artificial Intelligence}, 305:103662.

\bibitem[Ayawli et~al., 2019]{optimizedRRTstar}
Ayawli, B., Mei, X., Shen, M., Appiah, A., and Kyeremeh, E.~F. (2019).
\newblock Optimized rrt-a* path planning method for mobile robots in partially
  known environment.
\newblock {\em Information Technology And Control}, 48:179--194.

\bibitem[Cohen et~al., 2021]{ecbsct}
Cohen, L., Uras, T., Kumar, T. K.~S., and Koenig, S. (2021).
\newblock Optimal and bounded-suboptimal multi-agent motion planning.
\newblock In {\em Symposium on Combinatorial Search}.

\bibitem[Dubins, 1957]{1957}
Dubins, L.~E. (1957).
\newblock On curves of minimal length with a constraint on average curvature,
  and with prescribed initial and terminal positions and tangents.
\newblock {\em American Journal of Mathematics}, 79(3):497--516.

\bibitem[He et~al., 2016]{interpolatingBridges}
He, L., Pan, J., and Manocha, D. (2016).
\newblock Efficient multi-agent global navigation using interpolating bridges.
\newblock {\em CoRR}, abs/1607.07472.

\bibitem[Jiang and Wu, 2019]{memoryMARRT}
Jiang, J. and Wu, K. (2019).
\newblock Cooperative pathfinding based on memory-efficient multi-agent {RRT}.
\newblock {\em CoRR}, abs/1911.03927.

\bibitem[Karaman and Frazzoli, 2011]{RRTstar}
Karaman, S. and Frazzoli, E. (2011).
\newblock Sampling-based algorithms for optimal motion planning.
\newblock {\em CoRR}, abs/1105.1186.

\bibitem[Lan and Di~Cairano, 2015]{curvesRRT}
Lan, X. and Di~Cairano, S. (2015).
\newblock Continuous curvature path planning for semi-autonomous vehicle
  maneuvers using rrt.
\newblock In {\em 2015 European Control Conference (ECC)}, pages 2360--2365.

\bibitem[LaValle, 1998]{RRT98}
LaValle, S.~M. (1998).
\newblock Rapidly-exploring random trees : a new tool for path planning.
\newblock {\em The annual research report}.

\bibitem[Neto et~al., 2010]{SeventhOrderBezier}
Neto, A.~A., Macharet, D.~G., and Campos, M. F.~M. (2010).
\newblock Feasible rrt-based path planning using seventh order bézier curves.
\newblock In {\em 2010 IEEE/RSJ International Conference on Intelligent Robots
  and Systems}, pages 1445--1450.

\bibitem[Piegl and Tiller, 1996]{nurbs}
Piegl, L. and Tiller, W. (1996).
\newblock {\em The NURBS Book}.
\newblock Springer-Verlag, New York, NY, USA, second edition.

\bibitem[Russell and Norvig, 2020]{russel2010}
Russell, S. and Norvig, P. (2020).
\newblock {\em Artificial Intelligence: A Modern Approach}.
\newblock Prentice Hall.

\bibitem[Savla et~al., 2005]{DTSP}
Savla, K., Frazzoli, E., and Bullo, F. (2005).
\newblock On the point-to-point and traveling salesperson problems for dubins'
  vehicle.
\newblock volume~2, pages 786 -- 791 vol. 2.

\bibitem[Sederberg, 2012]{computer_aided_design}
Sederberg, T.~W. (2012).
\newblock {\em The NURBS Book}.
\newblock Brigham Young University.

\bibitem[Surynek, 2010]{psNPmapf}
Surynek, P. (2010).
\newblock An optimization variant of multi-robot path planning is intractable.
\newblock In {\em AAAI}.

\bibitem[Surynek, 2020]{SurynekSwarms}
Surynek, P. (2020).
\newblock Swarms of mobile agents: From discrete to continuous movements in
  multi-agent path finding.
\newblock In {\em 2020 IEEE International Conference on Systems, Man, and
  Cybernetics (SMC)}, pages 3006--3012.

\bibitem[Vana and Faigl, 2015]{faiglDTSP}
Vana, P. and Faigl, J. (2015).
\newblock On the dubins traveling salesman problem with neighborhoods.
\newblock In {\em 2015 {IEEE/RSJ} International Conference on Intelligent
  Robots and Systems, {IROS} 2015}, pages 4029--4034. {IEEE}.

\bibitem[\v{C}{\'{a}}p et~al., 2013]{MARRTstar}
\v{C}{\'{a}}p, M., Nov{\'{a}}k, P., Vok\v{r}{\'{\i}}nek, J., and
  P\v{e}chou\v{c}ek, M. (2013).
\newblock Multi-agent rrt*: Sampling-based cooperative pathfinding (extended
  abstract).
\newblock {\em CoRR}, abs/1302.2828.

\bibitem[\v{C}{\'{a}}p et~al., 2015]{COBRA}
\v{C}{\'{a}}p, M., Vok\v{r}{\'{\i}}nek, J., and Kleiner, A. (2015).
\newblock Complete decentralized method for on-line multi-robot trajectory
  planning in valid infrastructures.
\newblock {\em CoRR}, abs/1501.07704.

\bibitem[Verbari, 2017]{phdVerbari}
Verbari, P. (2017).
\newblock {\em Multi-RRT* : a sample-based algorithm for multi-agent planning}.
\newblock PhD thesis.

\bibitem[Walker et~al., 2018]{eicts}
Walker, T.~T., Sturtevant, N.~R., and Felner, A. (2018).
\newblock Extended increasing cost tree search for non-unit cost domains.
\newblock In {\em Proceedings of the 27th International Joint Conference on
  Artificial Intelligence, {IJCAI-18}}, pages 534--540. International Joint
  Conferences on Artificial Intelligence Organization.

\end{thebibliography}

\end{document}